%% file: Kompost_dilepton.tex
\documentclass[aps,prc,twocolumn,showpacs,preprintnumbers,superscriptaddress,
                          nofootinbib,float,longbibliography]{revtex4-1}

\usepackage{color}
\usepackage[dvipsnames]{xcolor}
\usepackage{graphicx}
\usepackage{dcolumn}
\usepackage{bm}
\usepackage{appendix}
\usepackage{multirow}
\usepackage[colorlinks=true, pdfstartview=FitV, linkcolor=blue,
citecolor=blue, urlcolor=blue]{hyperref}
\usepackage{mathrsfs}

\usepackage{xcolor}
\usepackage[normalem]{ulem}

\include{custom.tex}

\begin{document}


\title{Probing the equilibration of the QCD matter created in heavy-ion collisions\\with dileptons }

\author{Xiang-Yu Wu}
\affiliation{Department of Physics, McGill University, Montreal, QC, Canada H3A 2T8}
\affiliation{Department of Physics and Astronomy, Wayne State University, Detroit MI 48201}
\author{Lipei Du}
\affiliation{Department of Physics, McGill University, Montreal, QC, Canada H3A 2T8}
\affiliation{Department of Physics, University of California, Berkeley CA 94270}
\affiliation{Nuclear Science Division, Lawrence Berkeley National Laboratory, Berkeley CA 94270}
\author{Charles Gale}
\affiliation{Department of Physics, McGill University, Montreal, QC, Canada H3A 2T8}
\author{Sangyong Jeon}
\affiliation{Department of Physics, McGill University, Montreal, QC, Canada H3A 2T8}

\date{\today}

\begin{abstract}
A systematic study of intermediate invariant mass dilepton production in Pb+Pb collisions at $\sqrt{s_{NN}} = 5.02$ TeV is performed, using next-to-leading-order (NLO) thermal QCD dilepton emission rates with a multistage dynamical approach which includes event-by-event IP-Glasma initial conditions, relativistic viscous fluid dynamics, and a hadronic afterburner. Considering dilepton yield and anisotropic flow, special attention is paid to the out-of-equilibrium aspects, both thermal and chemical, and to the contribution of the Drell-Yan process. The relative contribution of each of those different channels to dilepton observables is calculated and discussed. 
\end{abstract}

\maketitle

\section{\label{sec:intro}Introduction}
Creating and measuring the properties of quark-gluon plasma (QGP), an exotic state of nuclear matter composed of colored quarks and gluons obeying quantum chromodynamics (QCD), is mainstream research in subatomic physics. This phase existed only microseconds after the big bang \cite{Jacak:2012dx}, could well constitute the inner core of massive stellar objects \cite{Annala:2023cwx}, and occupies a good fraction of the QCD phase diagram. To simulate the dense and high-temperature environment of the early universe in laboratory conditions, a practical approach is to collide two nuclei with extremely high collision energies, as is currently being performed at the Relativistic Heavy Ion Collider (RHIC) and the Large Hadron Collider (LHC).  An important milestone in the understanding and modeling of the QGP is the realization that the QGP fireball behaves like a fluid \cite{Gale:2013da} characterized by an extremely small specific shear viscosity \cite{Policastro:2001yc}. This is measured through correlations of final-state particles, more specifically by examining anisotropies in the azimuthal momentum  distribution of the final charged hadrons \cite{Poskanzer:1998yz}. 
Therefore, the  charged hadrons production and hadronic flow observables do carry information about the QGP medium. This degree of understanding has been achieved through years of developing a variety of theoretical approaches, evolving into what is currently known as multistage modeling. 

Whereas the information carried by hadrons is dominated by the conditions which exist at the freeze-out hypersurface, electromagnetic (EM) radiation is emitted throughout the space-time evolution and exits the hadronic medium.
This is true for photons \cite{Gale:2021emg,Paquet:2015lta,vanHees:2011vb} and for lepton pairs \cite{Gale:2003iz,Gale:1987ki,vanHees:2007th}. This class of observables is therefore complementary to that constituted by strongly interacting particles. 
The EM probes carry  information about the their space-time point of production, which is typically earlier than for hadronic production.
In view of those attributes, to calculate the yield of electromagnetic radiation it is essential to accurately simulate the dynamical  evolution history and to account for the different stages in a relativistic heavy ion collision, using a comprehensive hybrid framework. Microscopically, it is also crucial to precisely calculate the emission rate of EM probes. For instance, recent studies have considered modifying the emission rates of EM probes to include viscosity corrections \cite{Vujanovic:2019yih,Vujanovic:2016anq,Vujanovic:2017psb,Vujanovic:2013jpa}, corrections owing to the anisotropy of the emitting medium \cite{Hauksson:2017udm,Kasmaei:2018oag} or next-to-leading order (NLO) results at finite baryonic density \cite{Churchill:2023vpt}.

From the perspective of simulating the space-time evolution of the medium, the currently accepted minimum standard of multistage modeling \cite{Schenke:2010nt,Schenke:2010rr,Paquet:2015lta,Pang:2018zzo,Nijs:2020roc,Wu:2021fjf,Du:2023gnv} includes  fluctuating initial conditions, relativistic hydrodynamic evolution, and a hadronic afterburner. Models following this prescription have been used to predict and interpret a wide variety  of hadronic observables. Considering the different stages involved in this multistage model construction, the question of how ``hydrodynamization'' can be achieved in the very short time demanded by the interpretation of data ($\tau \lesssim 1$ fm/$c$) still remains somewhat mysterious, even though progress in the understanding of early time dynamics has been made in the last few years. 
For example, some effort has been devoted to studying the dynamical evolution of the preequilibrium (pre-hydrodynamics) stage, going from an initial state (either Glauber+free-streaming, or IP-Glasma \cite{Schenke:2012wb,Schenke:2012hg}) to viscous fluid dynamics. 
Alternatively, solving effective kinetic theories with anisotropic initial gluon distributions offers another approach to simulate the out-of-equilibrium stage \cite{Arnold:2002zm,Du:2020dvp,Garcia-Montero:2023lrd}, at the price of solving a set of coupled relativistic Boltzmann equations. However, recent work utilizing linear response theory offers a promising alternative to the brute force solution of the full non-linear Boltzmann equation, and propagates the energy-momentum tensor forward in time to be finally used as an initial state to the fluid-dynamical stage \cite{Kurkela:2018vqr,Kurkela:2018wud}. That model -- \kompost -- will be used later in this work to simulate the approach to kinetic equilibrium.

Concerning EM probes, the emission of dileptons is suppressed compared to that of real photons, owing to the presence of an extra factor $\alpha_{\rm em}$ in the production probability. However, an advantage of dileptons lies in an additional degree of freedom: the invariant mass $M$, independent of the reference frame. Recent studies  using a realistic space-time evolution model have shown a clear connection between the thermal dilepton invariant mass spectrum and the local temperature of the medium \cite{Churchill:2023zkk,Churchill:2023vpt,STAR:2024bpc}. 

In this work, we shall study the production of lepton pairs, including their invariant mass, transverse momentum, and anisotropic flow distributions,  
using the iEBE-MUSIC multistage model, following an approach very similar to that used for photons \cite{Gale:2021emg}. Importantly, and as was also done in Ref.~\cite{Gale:2021emg}, we will explore the ability of the dilepton signal to reveal details of the chemical composition of the medium from which the radiation is emitted. 

This paper is organized as follows: The next section briefly introduces the multistage model and details of the calculations of dilepton production and dilepton anisotropic flow from NLO emission rates. In Sec. \ref{sec:Results}, we first validate the model calibration using charged hadrons observables. We then explore in detail the effects of the preequilibrium stage and chemical equilibrium on dilepton production and dilepton anisotropic flow. 
A summary and outlook are provided in Sec. \ref{sec:conc}.

\section{\label{sec:Modeling} Theoretical framework}

\subsection{Multistage modeling}
In this work, the multistage modeling of heavy ion collisions is accomplished by the iEBE-MUSIC hydrodynamic approach \cite{Schenke:2010nt,Schenke:2010rr,Paquet:2015lta}, supplemented by IP-Glasma initial states \cite{Schenke:2012wb,Schenke:2012hg}, iSS particlization, and a UrQMD hadron afterburner \cite{Shen:2014vra,Bass:1998ca}. Since one of the goals of this study is to examine the contribution of the early-time dynamics, the  hybrid model  also incorporates  a nonequilibrium approach to the prehydrodynamic phase: \kompost.

Starting from the initialization, the primordial evolution of the system relies on IP-Glasma (the impact-parameter-dependent glasma), which samples nucleon positions from the Woods-Saxon distribution, with each nucleon modeled as three constituent quarks. Upon establishing the initial configuration of these quarks, the saturation scale $Q_s$ is determined via the IP-SAT model (impact-parameter-dependent dipole saturation model) \cite{Kowalski:2003hm}, which is calibrated using data from deeply inelastic scattering experiments at HERA \cite{Rezaeian:2012ji}. This saturation scale $Q_s$ correlates with the variance in the color charge density distribution. These serve as the sources for the small-$x$ classical gluon field within the color glass condensate (CGC) framework. The evolution of the gluon field is then described by the $SU(3)$ classical Yang-Mills (CYM) equations. Following the collision of two gluon fields, the field continues to evolve according to the CYM equations until the initial proper time $\tau_0^{\rm EKT}$ of the preequilibrium evolution\footnote{EKT stands for ``effective kinetic theory,'' the dynamical picture underlying \kompost.}. In this work, the initial proper time $\tau_0^{\rm EKT}$ is taken to be 0.1 fm/$c$ \cite{Kurkela:2018wud,Gale:2021emg}, which corresponds to a saturation scale $Q_s=2$ GeV.

At the initial proper time $\tau_0^{\rm EKT}$ of the preequilibrium phase, the complete energy-momentum tensor $T^{\mu\nu}_{\rm IPG}(\boldsymbol{x} )$ constructed from IP-Glasma constitutes the input of \kompost\!\!, as done in Ref.~\cite{Gale:2021emg}. 
For detailed discussions of \kompost\!, refer to \cite{Kurkela:2018wud,Kurkela:2018vqr}. In this study, the end of the preequilibrium phase or initial hydrodynamics time will be $\tau_{0}^{\rm hydro}$ = 0.8 fm/$c$, to be consistent with previous estimates \cite{Kurkela:2018wud,Gale:2021emg}. In the future, the proper time interval for preequilibrium could be determined through Bayesian inference \cite{Everett:2020xug,Heffernan:2023utr}. 

Following the preequilibrium evolution, the strongly interacting system evolves 
and is governed by relativistic hydrodynamics. In MUSIC, the initial energy density $\epsilon$ and initial flow velocity $u^{\mu}$ are initialized through Landau matching condition $u_{\mu}T^{\mu\nu}_{\rm EKT}= \epsilon u^{\nu}$, and the complete initial energy-momentum tensor is that passed on from \kompost\!. The dissipative parts of the energy-momentum tensor -- the shear tensor $\pi^{\mu\nu}$ and bulk pressure $\Pi$ -- are initialized  via
\begin{equation}
    \pi^{\mu\nu} = T^{\mu\nu}_{\rm EKT} - T^{\mu\nu}_{\rm ideal}(\epsilon), \quad \Pi = \frac{\epsilon}{3}- P(\epsilon),
\end{equation}
where $T^{\mu\nu}_{\rm ideal}(\epsilon)$ and $P(\epsilon)$ denote the ideal part of the energy-momentum tensor and pressure in the hydrodynamics phase, respectively\footnote{\kompost is conformal in its current formulation.}. Combined with the HotQCD Collaboration's lattice equation of state (EOS) \cite{HotQCD:2014kol}, the energy-momentum conservation and the second-order Israel-Stewart relaxation-type equations are solved numerically \cite{Schenke:2010nt}. In this study, the specific shear viscosity is taken as 0.12, as it is in the \kompost evolution. The bulk viscosity is modeled as temperature-dependent with a double Gaussian parametrization; for the details, please consult Ref.~\cite{Gale:2021emg}. The hydrodynamics evolution continues until the local energy density drops to $\epsilon_{\rm frz}$ = 0.18 GeV/fm$^3$. Subsequently, the Cooper-Frye prescription \cite{Cooper:1974mv} is used to convert the macroscopic thermodynamic variables of the hydrodynamics phase into the phase-space information of microscopic particles, via the iSS sampler module. During the sampling process, corrections for shear and bulk viscous components are applied according to the 14-moment method \cite{Ryu:2015vwa, Schenke:2020mbo}. Finally, the thermal hadrons are fed into the UrQMD model \cite{Shen:2014vra} to continue through hadronic cascading, including decay and formation channels.

\subsection{Dilepton production}
Considering three light flavors ($n_f = 3$), the differential thermal emission rate $d \Gamma_{l \bar{l}}$ 
of lepton pairs of energy $E$, three-momentum $\boldsymbol{P}$, and invariant mass $M =  \sqrt{E^2 - \boldsymbol{P}^2}$  is written as \cite{[{See, for example, }][{, and references therein.}]Kapusta:2023eix}

\begin{equation}
\frac{d \Gamma_{l \bar{l}}}{d^4 P}=\frac{2 \alpha_{\rm em}^2 f_B(E)}{9 \pi^3 M^2} B\left(\frac{m_l^2}{M^2}\right) \rho_V(E,\boldsymbol{P})\,,
\end{equation}
with the four-momentum of dilepton pair in the Minkowski coordinate in a given local fluid rest frame (LRF) 
\begin{equation}
P^{ \mu}=\left(M_{\perp} \cosh y, p_x, p_y, M_{\perp} \sinh y\right)\,,
\end{equation}
Here, $M_{\perp} = \sqrt{P_\perp^2 + M^2}$ is the transverse mass, $y$ the dilepton rapidity, and $f_B(E)$ the Bose-Einstein  distribution function. The kinematic factor for dilepton pair production is given by $B(x) = (1+2 x) \Theta(1-4 x) \sqrt{1-4 x}$, where $\Theta$ is the Heaviside step function. This work will focus on the production of electron-positron pairs, and their rest mass is neglected, leading to $B(0) = 1$. The vector spectral function, $\rho_V(E,\boldsymbol{P})$ used herein is derived using finite-temperature field theory, and is evaluated at NLO, including contributions from the Landau-Pomeranchuk-Migdal (LPM) effect \cite{Ghisoiu2014,Churchill:2023vpt}. The rates used here do not include viscous corrections; we leave those for future work.

Once the dilepton emission rate is determined, the dilepton production can be evaluated by integrating over the space-time volume $dV = \tau d \tau  d^2 \mathbf{x}_{\perp} d \eta_s $ which is provided by the dynamical evolution of  the QGP phase or preequilibrium phase. The differential dilepton spectrum is 
\begin{eqnarray}
\frac{d N_{l \bar{l}}}{M d M d y^{\prime}}=\int dV  \left.\frac{d \Gamma_{l \bar{l}}}{d^4 P}\right|_{P^{\mu} = \Lambda_\mu^\nu(U(X)) P^{\prime \nu}} p_T^{\prime} d p_T^{\prime} d \phi^{\prime}  
\label{eq:dilepton_yeild_M}\,,\\
\frac{d N_{l \bar{l}}}{p_T^{\prime} d p_T^{\prime} d y^{\prime}}=\int dV  \left.\frac{d \Gamma_{l \bar{l}}}{d^4 P}\right|_{P^{\mu} = \Lambda_\mu^\nu(U(X)) P^{\prime \nu}} M d M d \phi^{\prime} \,.
\label{eq:dilepton_yeild_pT}
\end{eqnarray}
As written, the differential emission rate $\frac{d \Gamma_{l \bar{l}}}{d^4 P}$ is evaluated in the LRF of the fluid cell. However,  dileptons are measured in the laboratory frame, which is denoted with a prime in  Eqs.~(\ref{eq:dilepton_yeild_M}) and (\ref{eq:dilepton_yeild_pT}). Therefore, when performing the above integration, the dilepton momentum in the laboratory frame, $P^{\prime \nu}$, must be boosted into the LRF of each flow cell using $P^\mu = \Lambda^\mu_\nu(U(X)) P^{\prime \nu}$, where $\Lambda^\mu_\nu$ is the Lorentz matrix which depends on the local flow velocity $U(X)$. For the sake of simplicity, the prime notation, which indicates the measured dilepton observables in the laboratory frame, and boost notation are omitted for the remainder of this paper.

Proceeding beyond the dilepton mass spectra, the $n$th order dilepton flow vector $v_n^{l \bar{l}}(\mathscr{X}) e^{i n \Psi^{l \bar{l}}(\mathscr{X})}$ can also be calculated from the fully differential emission rate in a single event, as defined in \cite{Paquet:2015lta,Vujanovic:2016anq,Vujanovic:2017psb}, specifically
\begin{eqnarray}
v_n^{l \bar{l}}(M) e^{i n \Psi^{l \bar{l}}(M)}=\frac{\int d V \frac{d \Gamma_{l \bar{l}}}{d^4 P} e^{i n \phi}  p_T d p_T d \phi d y}{\int d V \frac{d \Gamma_{l \bar{l}}}{d^4 P}  p_T d p_T d \phi d y}\,, \\
v_n^{l \bar{l}}(p_T) e^{i n \Psi^{l \bar{l}}(p_T)}=\frac{\int d V \frac{d \Gamma_{l \bar{l}}}{d^4 P} e^{i n \phi}  M d M d \phi d y}{\int d V \frac{d \Gamma_{l \bar{l}}}{d^4 P}  M d M d \phi d y} \,,
\end{eqnarray}
where $v_n^{l \bar{l}}$ and $\Psi^{l \bar{l}}$ are the anisotropic flow coefficient and event-plane angle of dilepton, respectively, and $\phi$ is the dilepton azimuthal angle.
The event-averaged dilepton flow coefficient can be evaluated using the scalar-product (SP) method by correlating the dilepton flow vector with the charged hadron flow vector $v_n^h$  \cite{Vujanovic:2019yih,Vujanovic:2016anq,Gale:2021emg,Vujanovic:2017psb,Paquet:2015lta}:
\begin{equation}
v_n^{l \bar{l}}\{{\rm SP}\}\left(\mathscr{X}\right)=\frac{\left\langle v_n^{l \bar{l}}\left(\mathscr{X}\right) v_n^h \cos \left\{n\left[\Psi_n^{l \bar{l}}\left(\mathscr{X}\right)-\Psi_n^h\right]\right\}\right\rangle}{\sqrt{\left\langle\left(v_n^h\right)^2\right\rangle}}.
\label{eq:vnsp}
\end{equation}
Here $\mathscr{X}$ could refer to either the invariant mass $M$ or transverse momentum $p_T$. In this study, the flow coefficient $v^h_n$ and event-plane angle $\Psi_n^h$ of charged hadrons are calculated with the kinematic cut $0.5<p_T<3.0$ GeV and $|\eta|<0.5$, and $\langle \cdot \rangle$ is the average over events.

\section{\label{results}Results}
\label{sec:Results}

\begin{figure}[htb]
	\centering
	\includegraphics[width=0.8\linewidth]{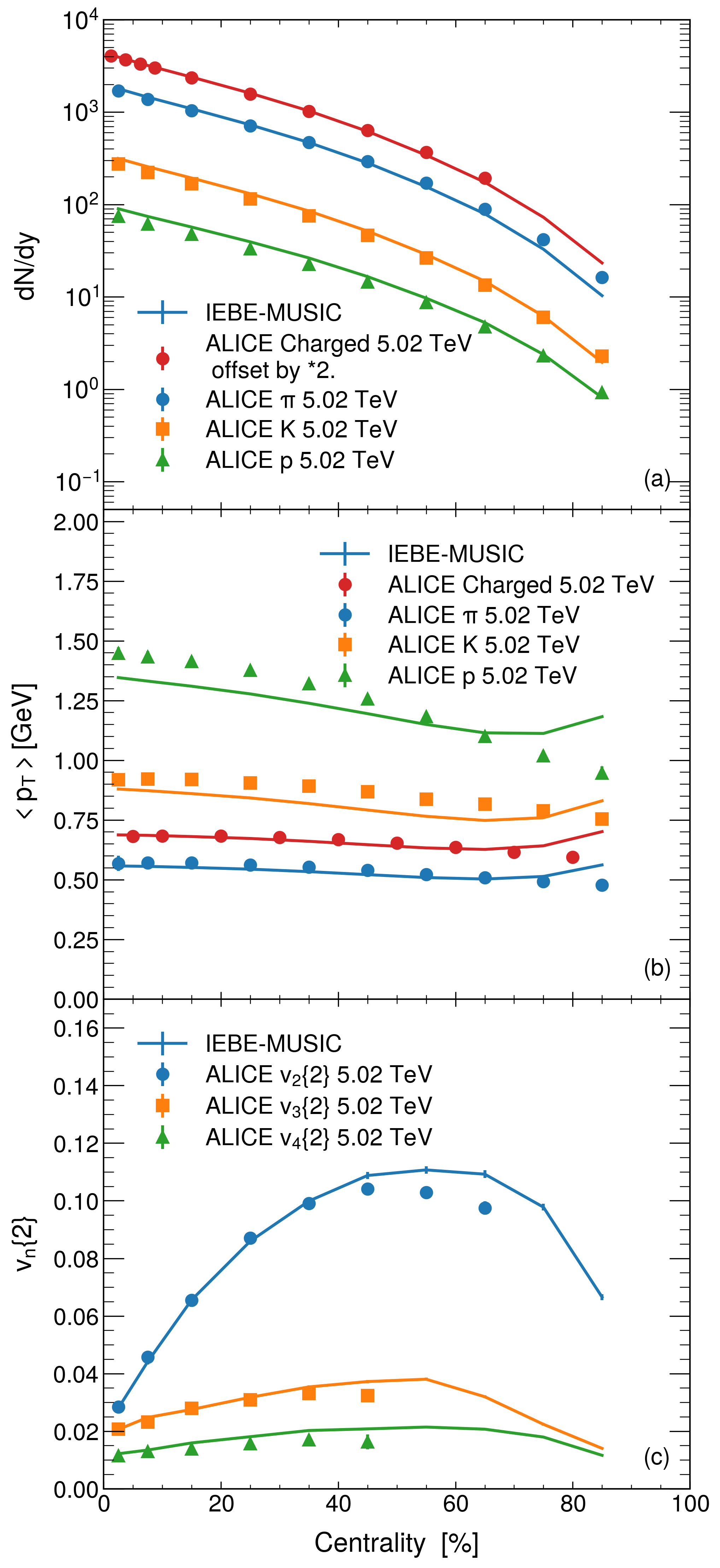}
	
 \caption{ The (a) multiplicity, (b) average transverse momentum and (c) momentum anisotropy coefficients $v_2\{2\}$ as a function of collision centrality for identical particle and charged hadrons in Pb+Pb collisions at the LHC, for a $\sqrt{s_{NN}}$ = 5.02 TeV collision energy. Note that (c) pertains only to charged hadrons.  The data were taken by the ALICE Collaboration \cite{ALICE:2019hno,ALICE:2016ccg}. }
	\label{fig:dNdy_pt_flow}
\end{figure}

We shall concentrate on the analysis of dilepton production and anisotropic flow in Pb+Pb collisions at a collision energy of $\sqrt{s_{NN}} = 5.02$ TeV. After our model is calibrated using hadronic flow observables, we will perform an analysis of the effects of chemical equilibration on dilepton observables. The dependence on both transverse momentum and invariant mass of dileptons will be considered. 

\subsection{Hadrons}

In Fig.~\ref{fig:dNdy_pt_flow}(a), the hadron multiplicity as a function of centrality bins in the central rapidity region for $\sqrt{s_{NN}} = 5.02$ TeV Pb+Pb collisions is shown. The experimental data are from the ALICE Collaboration \cite{ALICE:2019hno,ALICE:2016ccg}. The multiplicity is used to set the overall normalization of the initial energy-momentum tensor. It can be seen that the centrality dependence of the data for both charged particles and for identified hadrons is well reproduced. Note that the centrality selection in our study is determined from the final charged hadrons within the pseudorapidity range $|\eta| < 0.5$. This selection method is similar to what is used in the experimental analysis, allowing for an appropriate comparison between the model results and data.

In Fig.~\ref{fig:dNdy_pt_flow}(b), the mean transverse momentum $\langle p_T \rangle$ of identified hadrons is shown as a function of centrality bins, together with that of all charged hadrons. It can be observed that our results agree well with ALICE data for mesons (and also for all charged particles) up to the $\approx$75\% centrality bin. For protons produced in central collisions, the average transverse momentum is slightly underestimated ($\approx 7\%$) as was found in a recent similar study \cite{Schenke:2020mbo}. As was concluded there, future work will address this discrepancy; it is known that protons are especially sensitive to the details of the bulk viscosity \cite{Ryu:2015vwa}.  
For peripheral centralities, results and data deviate, with the model rising above the data as the centrality grows. 
This is understood as follows: our current model includes both preequilibrium (\kompost\!\!) and hydrodynamic (MUSIC)  evolutions, assuming a constant lifetime for the preequilibrium stage ($\approx$ 0.7 fm/$c$) across all centrality bins. In central collisions, the  lifetime of the hydrodynamic stage is significantly longer [$O(10)$ fm/$c$] than that of the preequilibrium stage. 
However, in peripheral collisions, the lifetime  of the hydro stage becomes smaller. 
As \kompost and IP-Glasma  both assume that the collision system has conformal symmetry \cite{Kurkela:2018wud} and thus vanishing bulk viscosity, the hydro phase has a comparatively shorter time to slow down the transverse expansion as compared with the situation in central collisions. 
In future studies, the lifetime of the preequilibrium stage could be considered as dependent on the initial energy density and centrality class, and could be determined via Bayesian inference based on experimental data \cite{JETSCAPE:2020mzn}.

In Fig. \ref{fig:dNdy_pt_flow}(c), the centrality dependence of different order anisotropic flows for final charged hadrons is presented. The flow coefficients $v_n\{2\}$ are calculated using the two-particle correlation method, with pseudorapidity $|\eta| < 0.5$ and transverse momentum $0.2 < p_T < 3.0$ GeV. Calculation results describe the experimental data up to the intermediate centrality bins but begin to deviate from the data beyond 45\% centrality, due to the relatively large contribution and deficiency of bulk viscosity in the early-time dynamics. Finally combined with the multiplicity, mean transverse momentum $\langle p_T \rangle$, and flow coefficient $v_n\{2\}$ of hadrons, it can be concluded that different hadronic observables present varying sensitivities to preequilibrium stages, but that the global representation of hadronic observables is more than adequate for our purposes. 

\subsection{Dileptons}

We estimate the dilepton production in the preequilibrium stage by using the same emission rate as in the hydrodynamic stage. The local ``temperature'' and fluid velocity can be obtained through the Landau matching method for the energy-momentum tensor of the preequilibrium stage with the HotQCD EOS. Thus, using the equilibrium emission rate is an effective method to estimate dilepton production of the preequilibrium stage. In addition, using the same EOS has the advantage of ensuring a smooth connection between the preequilibrium and hydrodynamic stages \cite{Gale:2021emg}. 
Note that the combination of (2+1)-dimensional hydrodynamics with the preequilibrium simulation is utilized to estimate the evolution of medium background at spatial rapidity $\eta_s$=0. However, dileptons may be emitted from regions with finite spatial rapidity \cite{Paquet:2022wgu}. Therefore, in Eqs.~(\ref{eq:dilepton_yeild_M}) and (\ref{eq:dilepton_yeild_pT}), spatial rapidity integration is performed by assuming the local temperature $T(x)$ and local flow velocity $U(X)$ to be independent of spatial rapidity within a narrow window, selecting $|\eta_s|<3$ in this work to evaluate observables at $y=0$. 

In the IP-Glasma model and \kompost model, both models assume that the collision system is purely gluonic. However, dilepton production originates from quark-antiquark annihilation at LO or, e.g., Compton scattering at NLO. This demands the presence of quarks and antiquarks in the collision system. Naturally, starting from all gluons, quarks and antiquarks will gradually be produced via the $gg\rightarrow q\bar{q}$ process, leading to the establishment of chemical equilibrium. Previous studies of chemical equilibration in heavy ion collisions have been performed with  EKT, and have defined a chemical relaxation time, $\tau_R = 4 \pi \eta/Ts$ \cite{Kurkela:2018xxd,Kurkela:2018oqw}. We adopt this definition here and model an effective suppression factor (SF):
\begin{equation}
{\rm SF}(T,\tau) = 1-e^{- A \frac{\tau}{\tau_R (T)}}
\label{SF}
\end{equation}
Inspired by the work in Refs. \cite{Kurkela:2018oqw,Kurkela:2018xxd}, where the time for chemical equilibrium is found to be $\tau_{\rm chem} = 1.2\, \tau_R$, the constant $A$ in Eq.~(\ref{SF}) is determined by requiring ${\rm SF}(\tau_{\rm chem}) = 0.9$, as in those references. This suppression factor multiplies the dilepton emission rate and represents the fact that the fermions are under-populated as compared to the situation in chemical equilibrium. Since our rates are derived up to, and including NLO, we allow for flexibility by studying cases where the emission is either linear or quadratic in the SF, in addition to the case where SF=1.  Note that the suppression factor thusly generated resembles that in Ref.~\cite{Gale:2021emg} -- a study similar to that performed here, but for photons -- with the difference that the approach to equilibrium is temperature dependent here. 

Obtaining the fermionic content in the non-equilibrium environment from first principles would be eminently desirable. However, while this is possible in principle, it would entail developing a multistage model where a non-equilibrium transport dovetails to a viscous hydro, and where all parameters would be tuned (perhaps through a Bayesian study) to observables; this is the topic of future work. Instead, we use here this flexible suppression factor to explore the sensitivity of the dilepton signal to its specific form.

\begin{figure}[htb]
	\centering
	\includegraphics[width=0.9\linewidth]{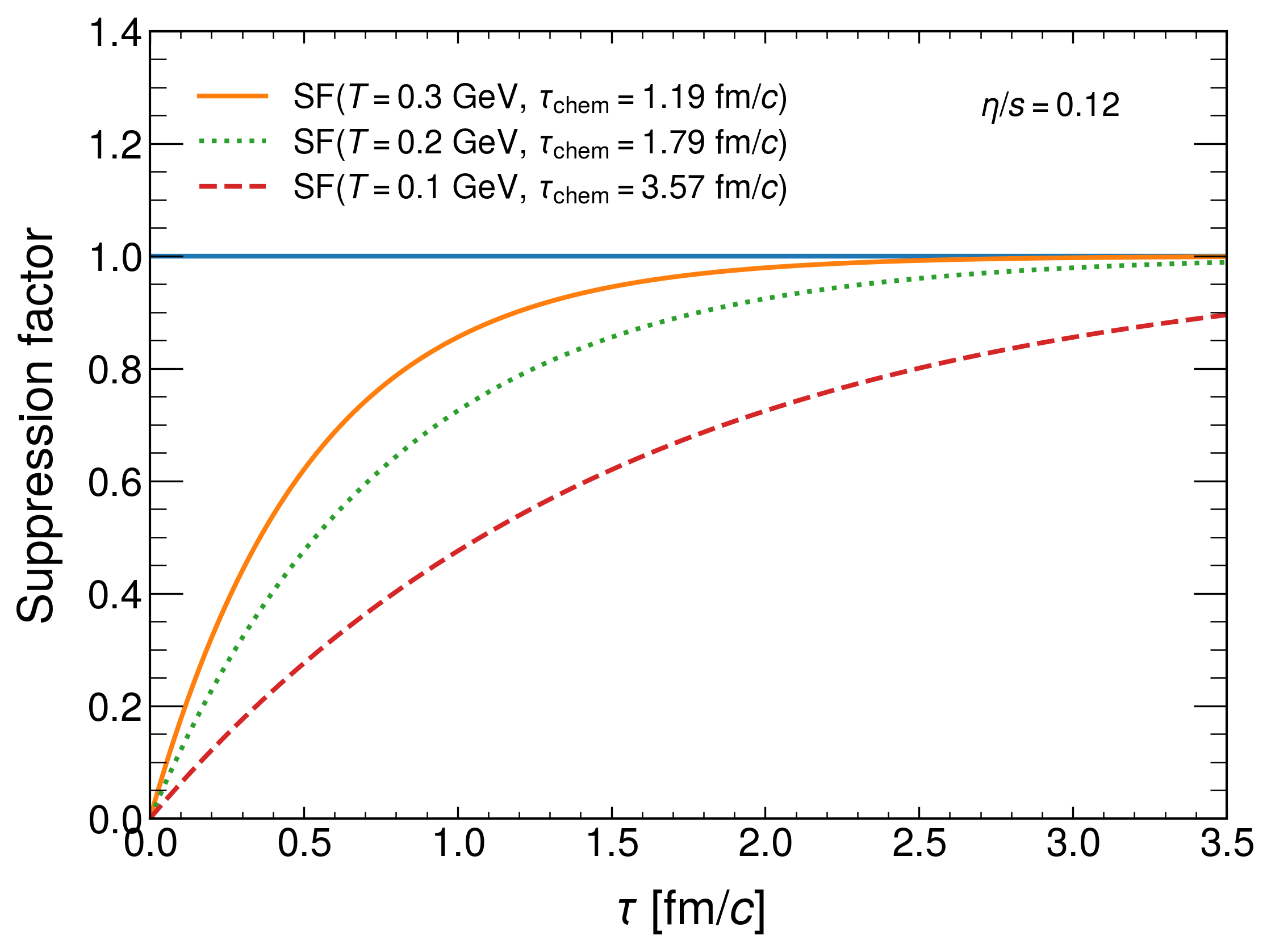}
	 \caption{The suppression factor, plotted as a function of proper time, simulates the effective chemical equilibrium in preequilibrium stage at three typical local temperatures: 0.3 GeV (solid line), 0.2 GeV (dotted line), and 0.1 GeV (dashed line). In this figure, $\tau$ is the elapsed proper time measured after the beginning of the \kompost\ phase.}
	\label{fig:sp_factor}
\end{figure}

In Fig.~\ref{fig:sp_factor}, the suppression factor is presented as a function of proper time for three typical local temperatures: $T$ = 0.1, 0.2, and 0.3 GeV, corresponding to chemical equilibrium times of 1.19 , 1.79 , and 3.57 fm, respectively. The shear viscosity per unit entropy density is set at 0.12, consistent with our simulation of the preequilibrium and hydrodynamic stages. From the Fig.~\ref{fig:sp_factor}, it is evident that systems with higher temperatures reach chemical equilibrium faster than those at lower temperatures. The time of chemical equilibrium may be slightly longer than the initial proper time of hydrodynamic evolution \cite{Kurkela:2018xxd}. We will now examine the details of the production of lepton pair using our multistage model, with a dynamical preequilibrium stage. 

\subsubsection{M-dependent observables}

\begin{figure}[htb]
	\centering
	\includegraphics[width=0.9\linewidth]{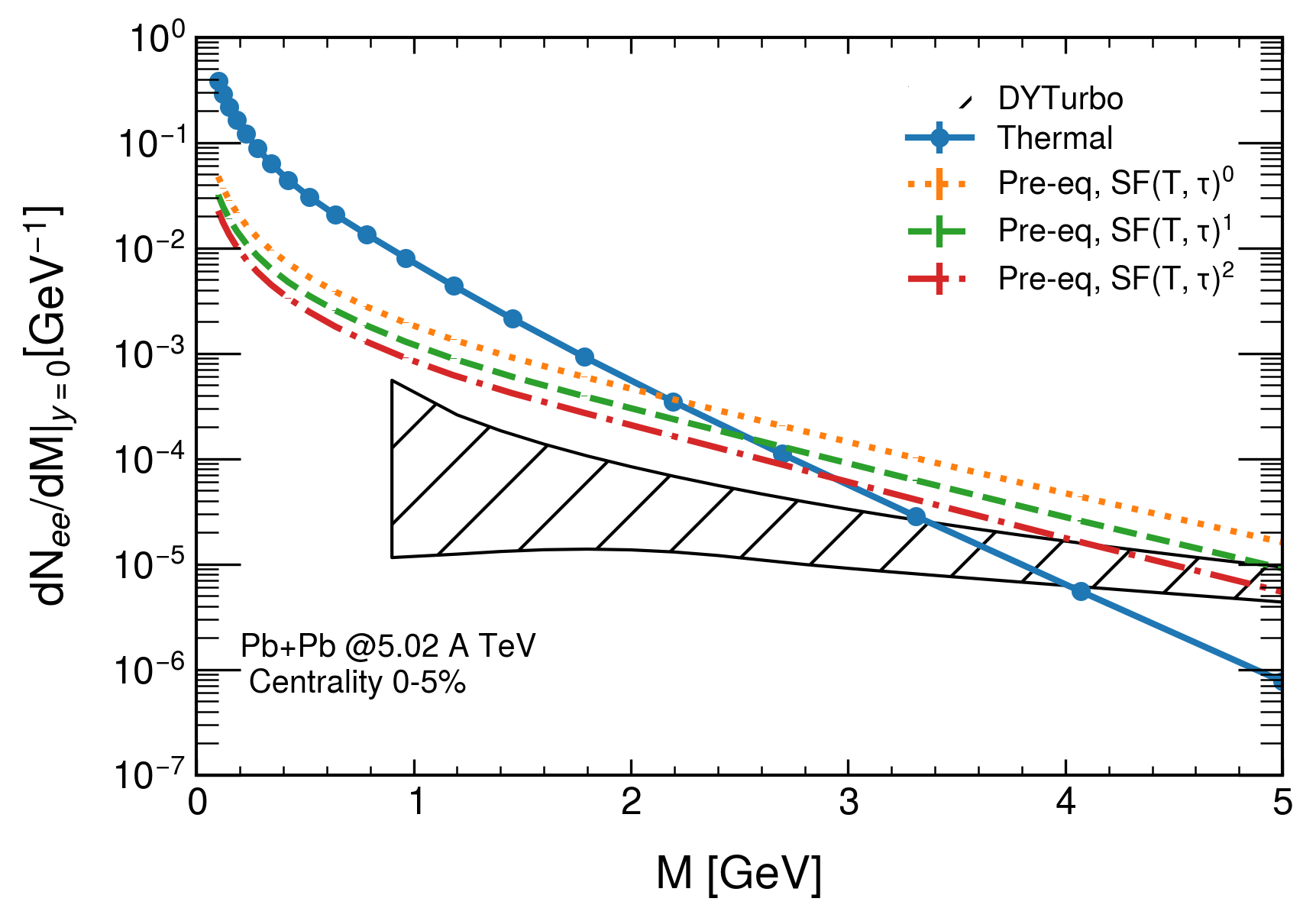}
	
 \caption{ The yield of dileptons, induced by various sources, as a function of invariant mass $M$ in the 0–5\% centrality at $\sqrt{s_{NN}}$ = 5.02 TeV  Pb + Pb collisions at the LHC . The band shows results of the DYTurbo package for Drell-Yan calculation, and the spread reflects the scale uncertainty discussed in the text. Here and elsewhere, the curves labeled with powers of SF correspond to the cases with no suppression (exponent 0), with the dilepton rates linear in the suppression factor defined in the text (exponent 1), and with a quadratic suppression factor (exponent 2). }
	\label{fig:th_pr_DY}
\end{figure}
\begin{figure}[htb]
	\centering
	\includegraphics[width=0.77\linewidth]{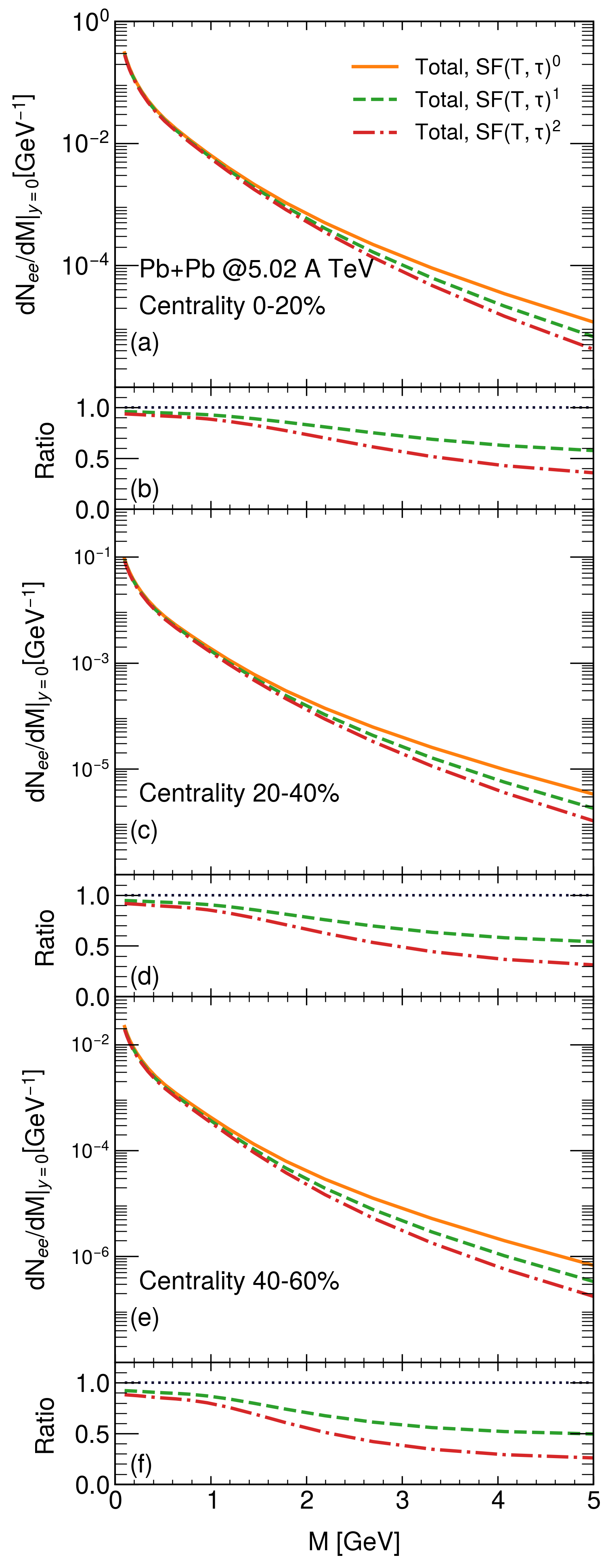}
	
 \caption{The production of dileptons as a function of invariant mass in Pb+Pb collisions at LHC energies, across centrality classes of 0-20\%, 20-40\%, and 40-60\%, is calculated considering instant chemical equilibration (SF)$^0$, or a linear or quadratic power of the suppression factor. The contributions shown (``Total'') sum those from the preequilibrium and hydro stages. Panels (b), (d), and (f) show the ratio of the yields with respect to that in chemical equilibrium. }
	\label{fig:dNdM_centrality}
\end{figure}
\begin{figure}[htb]
	\centering
	\includegraphics[width=0.8\linewidth]{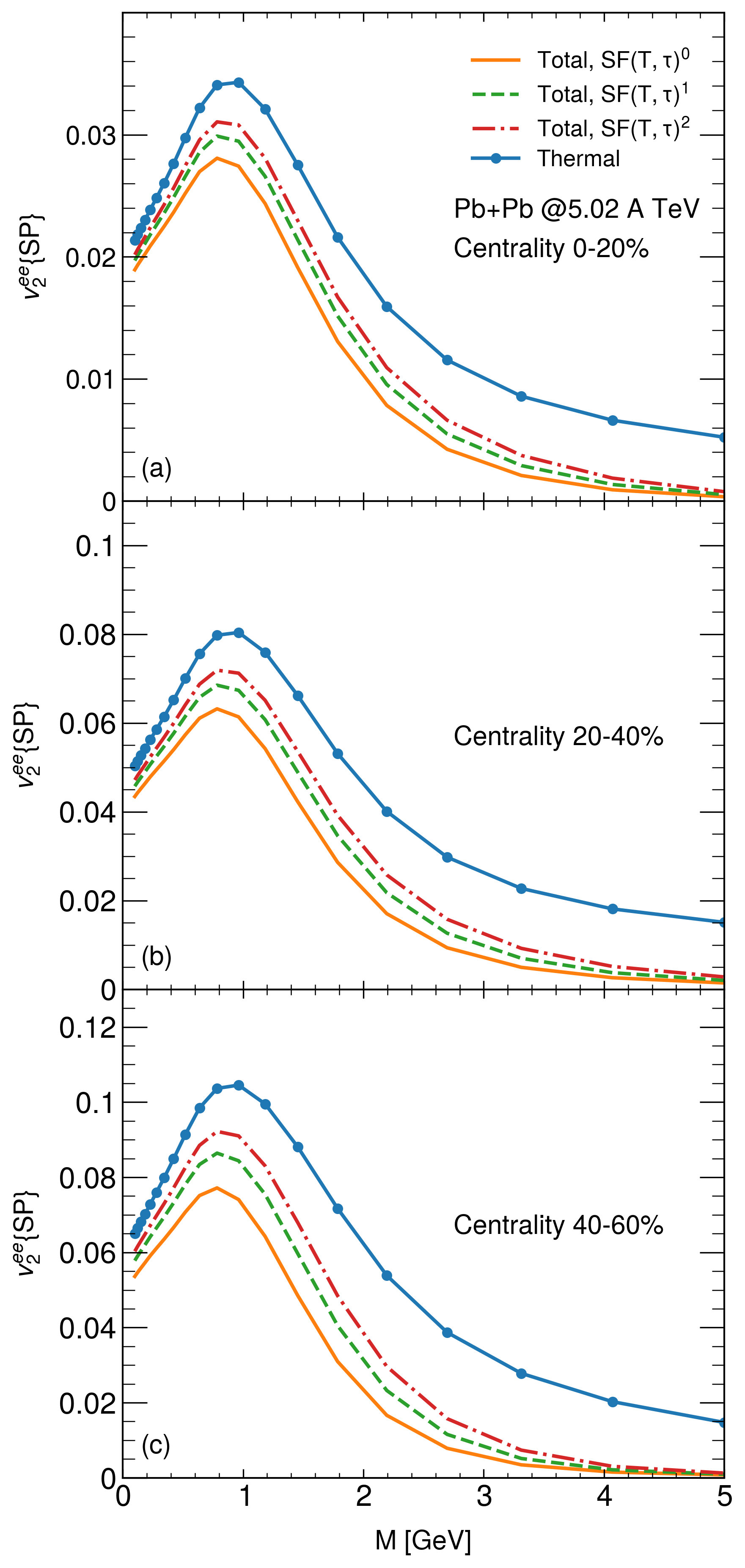}
	
 \caption{The dilepton elliptic flow $v^{ee}_2(M)$  as a function of invariant mass in Pb+Pb collisions at $\sqrt{s_{NN}}$ = 5.02 TeV collision energy, for centrality classes of 0-20\%, 20-40\%, and 40-60\%. The lines represent sources of the dilepton elliptic flow $v^{ee}_2(M)$: from the hydrodynamics stage (circle line), and the combined results (``total'') from both the hydrodynamics and preequilibrium stages, with instant chemical equilibrium (solid line), or assuming a linear (dashed line), or quadratic (dot-dashed line) power of the suppression factor.  }
	\label{fig:v2_M_centrality}
\end{figure}

\begin{figure}[htb]
	\centering
	\includegraphics[width=0.8\linewidth]{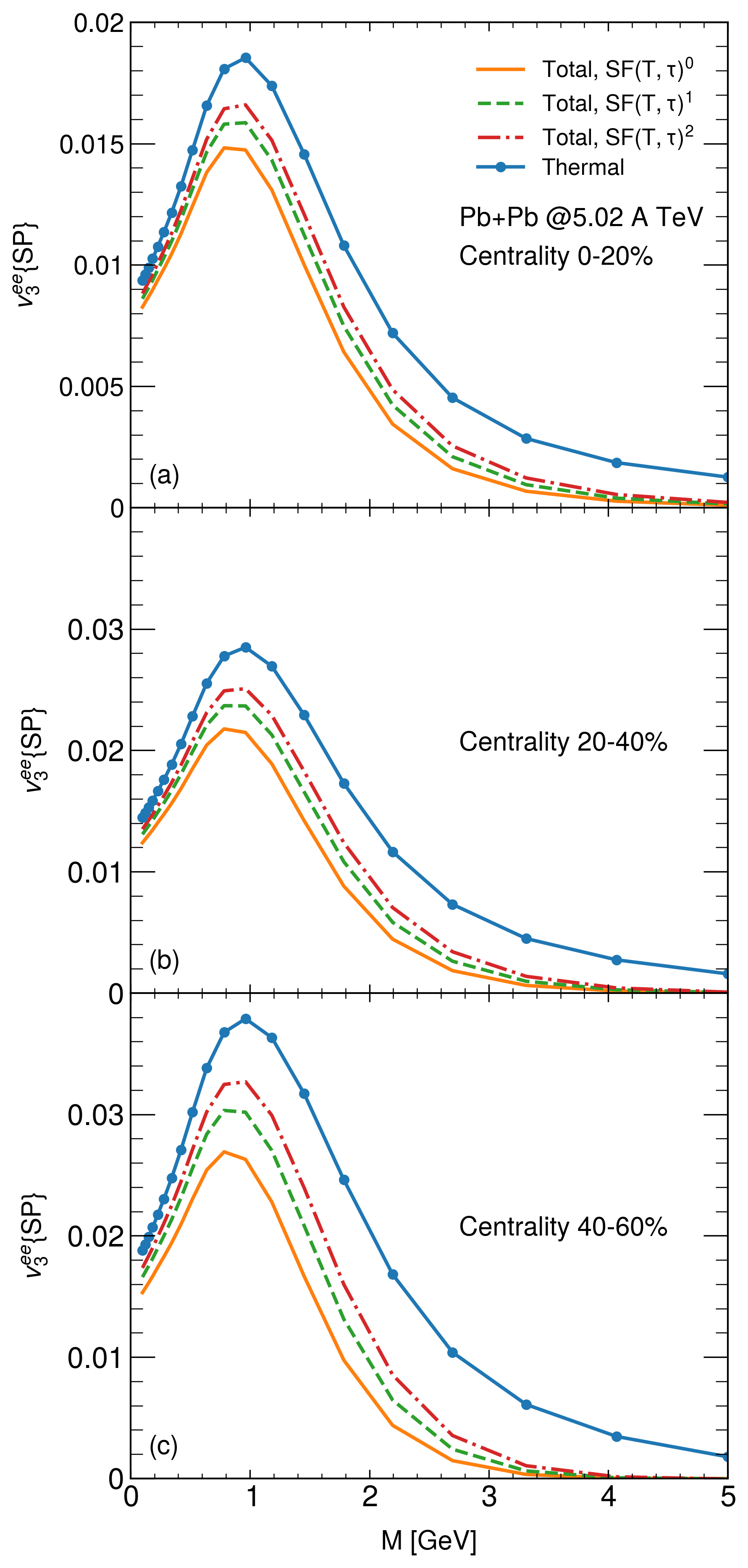}
	
 \caption{As in Fig.\ref{fig:v2_M_centrality}, but for the dilepton triangular flow $v^{ee}_3(M)$.  }
	\label{fig:v3_M_centrality}
\end{figure}

\begin{figure*}[htb]
	\centering
	\includegraphics[width=0.7\linewidth]{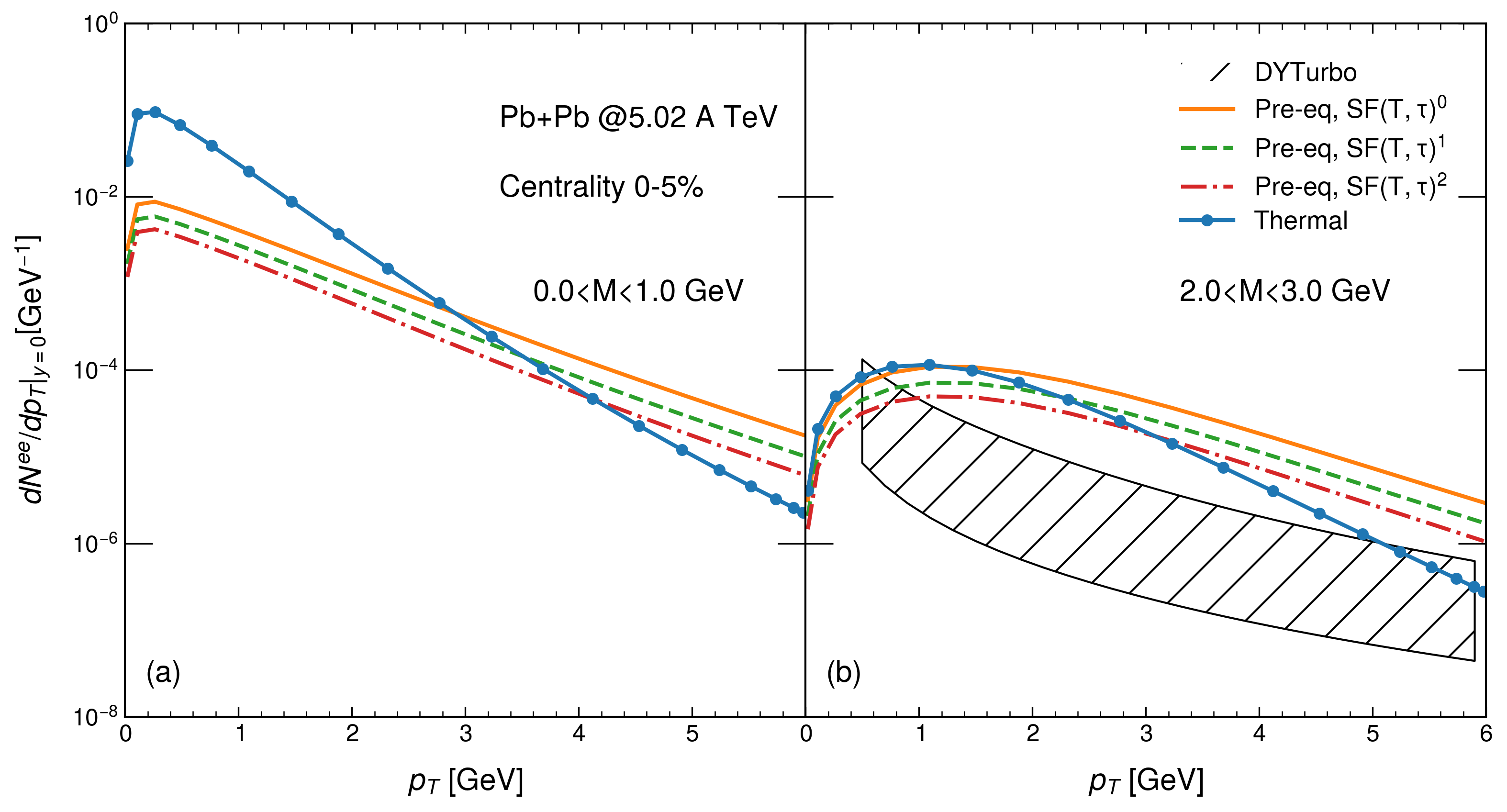}
	
 \caption{ The thermal dilepton production and preequilibrium dilepton production under varying orders of chemical equilibrium as a function of transverse momentum ($p_T$) across different invariant mass windows: (a) $0.0 < M < 1.0$ GeV and (b) $2.0 < M < 3.0$ GeV in 0-5\% centrality class Pb+Pb collision with $\sqrt{s_{NN}}=5.02$ TeV.}
	\label{fig:dNdpt_M_0_5}
\end{figure*}

\begin{figure*}[htb]
	\centering
	\includegraphics[width=0.7\linewidth]{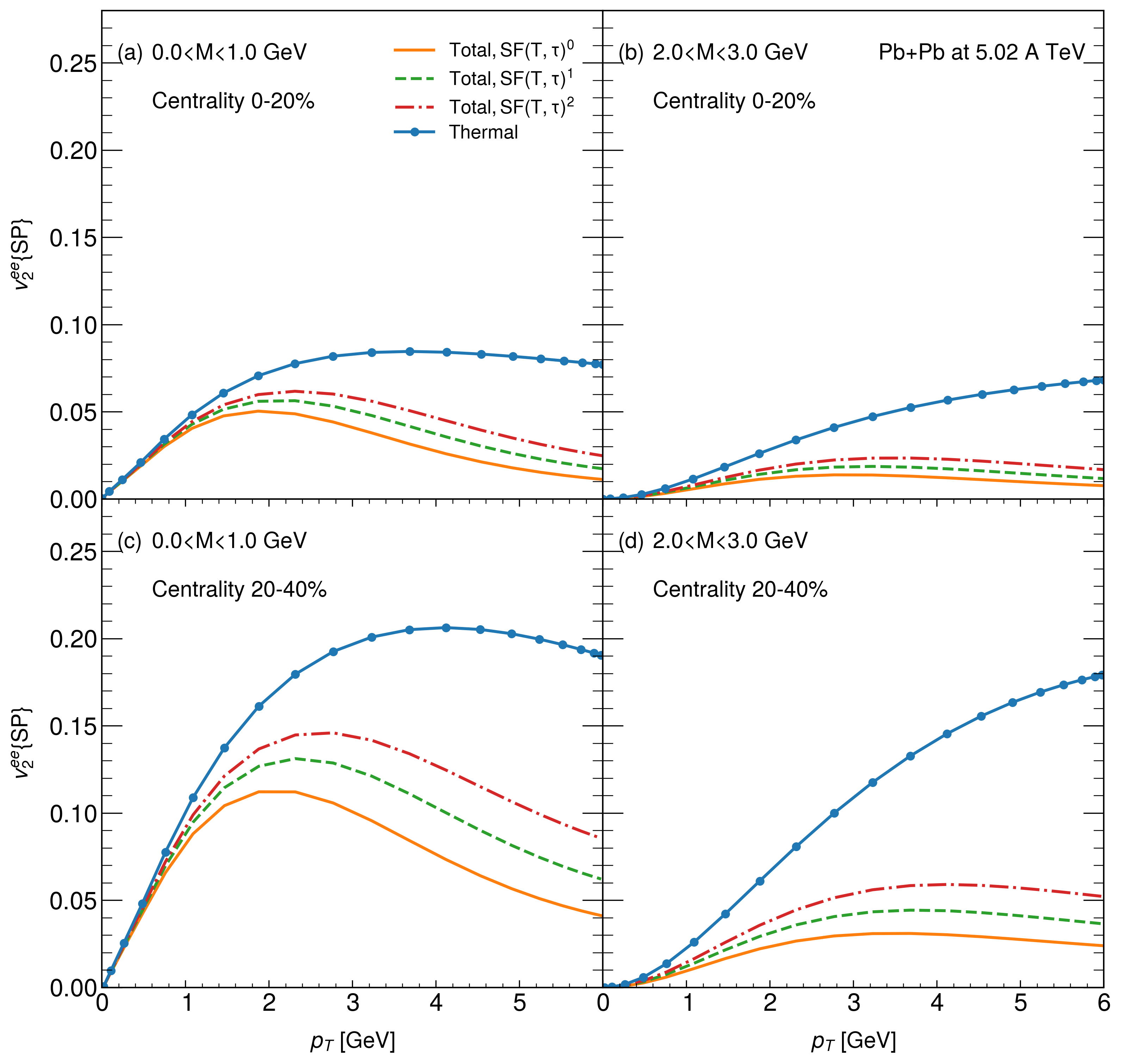}
	
 \caption{ Thermal dilepton and total elliptic flow $v^{ee}_2(p_T)$  as functions of transverse momentum $p_T$ across different invariant mass windows in Pb + Pb collisions at the LHC. Panels (a) and (c) show results for  $0.0 < M < 1.0$ GeV in the 0-20\% and 20-40\% centrality classes, respectively. Panels (b) and (d) correspond to $2.0 < M < 3.0$ GeV  for the 0-20\% and 20-40\% centrality classes. The varying orders of chemical equilibrium are considered in total elliptic flow.}
	\label{fig:v2_pt_centrality}
\end{figure*}

The research in this paper focuses mainly on thermal dileptons. Therefore, for the calculations of dilepton production and dilepton flow observables, only cells where the local temperature $T(X)$ exceeds the chemical freeze-out temperature [$T(X)>T_{\rm frz}=0.166$  GeV] \cite{STAR:2017sal} are considered.

In Fig.~\ref{fig:th_pr_DY}, the dilepton yield is presented as a function of invariant mass in the 0-5\% centrality class at $\sqrt{s_{NN}} = 5.02$ TeV Pb+Pb collisions at the LHC. The dilepton production contains different contributions including the preequilibrium stage -- which is analyzed under three kinds of chemical equilibrium scenarios -- the hydrodynamics stage (labeled as thermal), and the Drell-Yan process \cite{Drell:1970wh}. 
At LHC collision energies, the Drell-Yan process is one of the main sources of dilepton production at high invariant masses. It is an important test of the standard model, and has been measured for high mass dileptons in p+p collisions at the LHC \cite{ATLAS:2014alx}. However, in the low and intermediate invariant mass regions in heavy-ion collisions, both measurements and theoretical calculations of the Drell-Yan process are scarce.

\begin{figure*}[htb]
	\centering
	\includegraphics[width=0.7\linewidth]{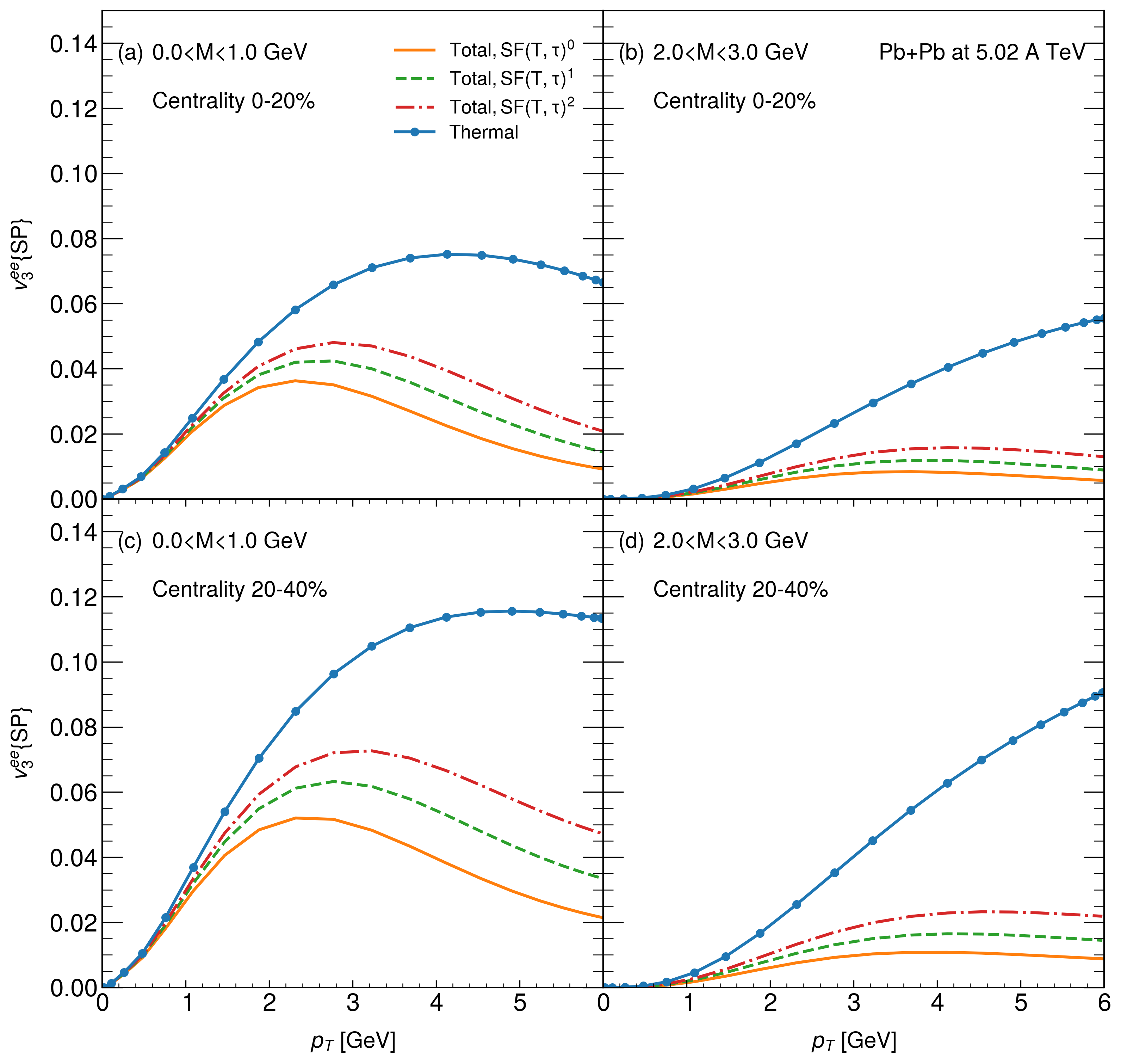}
 \caption{  As in Fig.~\ref{fig:v2_pt_centrality}, but for $v_3^{ee}(p_T)$.}
	\label{fig:v3_pt_centrality}
\end{figure*}

In this work we utilize the DYTurbo package \cite{Camarda:2019zyx} to estimate the production of dileptons from the Drell-Yan process at low and intermediate invariant masses within a rapidity window $|y|<1$. The DYTurbo program is designed to calculate the Drell-Yan process at NLO in p+p collisions, so it requires adjustments for A+A collisions. These adjustments include considering the nuclear parton distribution functions nPDFs (EPPS16nlo-CT14nlo-Pb208 \cite{Eskola:2016oht}), as well as accounting for the number of binary collisions $N_{coll}$ as determined by the Monte Carlo Glauber model\cite{Loizides:2017ack}, and the inelastic cross section $\sigma_{in}^{pp}$ from p+p collisions, in the cross-section calculations performed by DYTurbo:
\begin{equation}
\frac{d N_{e e}^{D Y}}{d M d y}=\frac{d \sigma_{e e}^{D Y;pp} }{d M d y} \cdot \frac{N_{coll}}{\sigma_{in}^{pp}}
\end{equation}

In  Fig.~\ref{fig:th_pr_DY}, the band obtained with DYTurbo representing dilepton production from Drell-Yan  arises from the variation in the factorization and renormalization scales. The upper limit corresponds to the factorization and renormalization scales being twice the invariant mass of the lepton pair, while the lower limit corresponds to half the mass \cite{Coquet:2021lca}. In perturbative QCD calculations, parton distribution functions and fragmentation functions are typically limited from below by a practical scale of the order of  $Q \approx 1$ GeV. This is also the case for DYTurbo and this explains the absence of results at lower invariant masses. Note that in addition to using DYTurbo to estimate dilepton production from the Drell-Yan process, the experimental community also often relies on Pythia6 \cite{Sjostrand:2006za,STAR:2015tnn,STAR:2015zal}. Results obtained with DYTurbo and Pythia6 will be compared in the Appendix~\ref{sec:appendix}. 

In Fig.~\ref{fig:th_pr_DY}, it is observed that thermal dilepton production dominates in the low invariant mass range compared to other dilepton sources. This is because dileptons at low invariant mass are primarily produced in relatively low-temperature regions, which correspond to the later stages of heavy-ion collisions. As the invariant mass increases, the contribution of dileptons from the preequilibrium stage gradually increases and becomes the dominant source over thermal dileptons for \(M > 2.2\) GeV. 
Keep in mind that dilepton production from the Drell-Yan process in the invariant mass region from 1 to 4 GeV exhibits large theoretical systematic uncertainty, even though the Drell-Yan process can be accurately measured at high invariant mass (\(M > 66\) GeV) \cite{ATLAS:2014alx}. It is also found that dilepton production through  Drell-Yan is consistently smaller than the preequilibrium contribution until $M\gtrsim 4$ GeV. 
Taken together, those observations indicate that dilepton production from the preequilibrium stage is observable in the intermediate invariant mass region, and that any quantitative differences with previous results \cite{Kasmaei:2018oag,Churchill:2020uvk,Coquet:2021lca,Garcia-Montero:2024lbl} highlight the importance of realistic initial states, of complete emission rates,  and of complete modeling. Regarding the effect of the chemical equilibrium, it is observed that dileptons induced from the preequilibrium phase with a higher-power of the SF chemical equilibrium factor will be suppressed. In addition, the higher mass regions directly correlates with high temperatures \cite{Churchill:2023vpt}.  Thus, dilepton production in the intermediate invariant mass region will be sensitive to the early chemical equilibrium, as shown in Fig.~\ref{fig:th_pr_DY}.

By summing over the contributions from the hydrodynamics and preequilibrium stages first shown in Fig.~\ref{fig:th_pr_DY}, we plot the total thermal dilepton production as a function of invariant mass $M$ in the 0-20\%, 20-40\%, and 40-60\% centrality classes at LHC collision energy in panels (a), (c), and (e) of Fig.~\ref{fig:dNdM_centrality}. Each panel includes three different curves corresponding to three scenarios of chemical equilibrium. Panels (b), (d), and (f) present the ratios between the linear and quadratic orders to the instant chemical equilibrium. In each centrality class, the effect of chemical equilibrium primarily influences the intermediate invariant mass region. From the panels of ratios, it is clearer to see that higher orders of chemical equilibrium suppress dilepton production more significantly, especially at an invariant mass of 4-5 GeV, where only 40\% of dilepton production with quadratic order suppression factor is observed compared to the instant chemical equilibrium. Additionally, the effect of chemical equilibrium generally increases with the centrality class and the difference between instant and quadratic chemical equilibrium becomes more distinct in peripheral collisions. This can be attributed to the relative contribution of the preequilibrium stage which grows in peripheral collisions.

We now discuss the dilepton anisotropic flow coefficients.  In all of the flow calculations in this work, we do not include the contribution of Drell-Yan dileptons. Their elliptic flow can be assumed to vanish, and their yield can, however, be calculated, either with DYTurbo or Pythia (see the Appendix) and subtracted from measured signals, as done for the pQCD production of real photons  \cite{Esha:2023ooh}. Clearly, a direct measurement in p+p collisions would be advantageous, as it would largely alleviate the scale uncertainty found in DY calculations. Nevertheless, we provide the following estimates: for the invariant mass-dependent flow (Fig.~\ref{fig:v2_M_centrality}) at M = 2 GeV, including the DY contribution reduces the elliptic flow by an amount varying between 2\% and 8\% (depending on the choice of scales) in the 0-20\% centrality bin. In the 20-40\% and 40-60\% bins, the reduction ranges from 2\% to 11\% and from 3\% to 16\%, respectively.

Figures.~\ref{fig:v2_M_centrality} and \ref{fig:v3_M_centrality}  show the $p_T$-integrated thermal and total dilepton elliptic flow $v^{ee}_2(M)$ and triangular flow $v^{ee}_3(M)$, in the  0-20\%, 20-40\%, and 40-60\% centrality classes. The different order of chemical equilibrium (power of the SF) is shown in the total dilepton flow coefficient. In a given centrality class, the elliptic flow coefficient  $v^{ee}_2(M)$ initially increases, reaches a peak at about $M=1$GeV, then starts to decrease as the invariant mass $M$ increases, owing to the smaller momentum anisotropy of the collision system at the early stage. After considering the contribution of the preequilibrium stage, the total dilepton $v^{ee}_2(M)$ will be suppressed due to the almost zero momentum anisotropy in the preequilibrium stage. Particularly, the total dilepton flow at the invariant mass larger than 2 GeV region, where the preequilibrium is dominant, shows a more significant suppression. When focusing on the effect of chemical equilibrium, it displays a trend opposite to that seen in the dilepton yields. For the higher order of the suppression factor, the total dilepton flow is closer to that of thermal dileptons. This is because a higher-order suppression factor will further suppress the contribution of dilepton production at the early stage, so the relative contribution of later proper times in preequilibrium will be enhanced as the flow builds up as time grows.

Next, we examine the centrality dependence. Notably, the dilepton flow will grow as the centrality increases, owing to a more elliptical source in coordinate space. For the dilepton triangular flow $v^{ee}_3(M)$, a similar conclusion can be obtained. The dilepton triangular flow $v^{ee}_3(M)$ shows a weaker centrality dependence compared to the dilepton elliptic flow $v^{ee}_2(M)$, as it mostly originates from  event-by-event fluctuations. It will slightly increase in the peripheral collision due to the larger fluctuations in the QGP background.

\subsubsection{$p_T$-dependent observables}

In addition to studying the invariant mass dependence of dilepton production and of flow, it is equally revealing to examine their  transverse momentum ($p_T$) dependence because different invariant mass windows can provide insight into various periods of the QGP evolution history. Figure.~\ref{fig:dNdpt_M_0_5} displays the dilepton production as a function of transverse momentum $p_T$ in different invariant mass windows: (a) $0 < M < 1$ GeV and (b) $2 < M < 3$ GeV, within the 0-5\% centrality class in $\sqrt{s_{NN}}=5.02$ TeV Pb+Pb collisions. 
Within the invariant mass window of $0 < M < 1$ GeV, which predominantly originates from the low-temperature region of the QGP medium, the dilepton spectrum is consequently dominated by the thermal production. 
However, beginning in the intermediate $p_T$ region, dileptons from the preequilibrium stage become dominant. Additionally, it is observed that the slope of dilepton production in preequilibrium is flatter, reflecting the natural property for dileptons in preequilibrium to possess higher transverse momentum.
In the $2<M<3$ GeV invariant mass window, the preequilibrium dilepton production is seen to be comparable to the thermal dilepton production in the low $p_T$ region. Especially in the case of instant chemical equilibrium, the preequilibrium dilepton production consistently exceeds  thermal dilepton production and Drell-Yan dileptons. This indicates that by carefully selecting the invariant mass window it is possible to probe the properties of the preequilibrium stage using dilepton production. Furthermore, in this centrality class the transverse momentum dependence of dilepton production is only moderately influenced by the effects of chemical equilibrium. 

In Fig.~\ref{fig:v2_pt_centrality}, the dilepton flow $v^{ee}_2(p_T)$ is presented as a function of transverse momentum ($p_T$) within the invariant mass windows of $0 < M < 1$ GeV and $2 < M < 3$ GeV, for the 0-20\% and 20-40\% centrality classes. The total dilepton flow $v^{ee}_2(p_T)$ is also calculated assuming different values of the suppression factor, as explained previously.  In the low invariant mass range, shown in panels (a) and (c), which predominantly originates from the low-temperature region of the QGP medium, thermal dileptons produce a sizable elliptic flow. However, after accounting for preequilibrium dileptons, the total dilepton flow is significantly suppressed relative to the thermal dilepton flow, particularly at higher transverse momenta $p_T$, where preequilibrium dileptons are known to dominate, as demonstrated in Fig.~\ref{fig:dNdpt_M_0_5}.
In the intermediate invariant mass range [panels (b) and (d)], it is even more obvious that the total dilepton elliptic flow experiences a noticeable  reduction due to the dominance of preequilibrium dileptons in this invariant mass range. Furthermore, it is interesting to note that the dilepton total elliptic flow is very sensitive to the effects of chemical equilibrium. The ratios of total dilepton flow with a linear and quadratic SF to values assuming instant chemical equilibrium are approximately 1.5 and 2.5, respectively, in the 20-40\% centrality class of panel (c). This 
indicates that, considering all centrality classes, dilepton flow does provide a good potential for quantifying the effects of chemical equilibrium in the preequilibrium stage of heavy ion collisions. The same conclusion can also be obtained from the analysis of dilepton triangular flow $v^{ee}_3(p_T)$ in Fig.~\ref{fig:v3_pt_centrality}.

\section{Conclusion}\label{sec:conc}

In this work, an analysis of thermal dilepton production and anisotropic flow is performed in Pb+Pb collisions at a LHC collision energy of \(\sqrt{s_{NN}} = 5.02\) TeV. A state-of-the-art multistage model, comprising the QCD-based IP-Glasma, the out-of-equilibrium  \kompost, the relativistic hydrodynamic MUSIC, and the hadronic afterburner UrQMD, is used to simulate the entire dynamic evolution of relativistic heavy ion collisions.
After calibrating the model, we investigate dilepton production at the different stages, using dilepton production rates complete at NLO in the strong coupling. It is found that thermal dileptons are the dominant source at low invariant mass, and preequilibrium dileptons overtake thermal dileptons to become the main contribution at intermediate invariant mass ranges, exceeding the dilepton production from the Drell-Yan process. \kompost was used to model the approach to kinetic equilibrium, and the effect of chemical equilibrium has been incorporated therein. We find that the departure from chemical equilibrium can suppress dilepton production and enhance dilepton anisotropic flow. 

This paper also studies the transverse momentum dependence of dilepton observables. One interesting result is that the transverse momentum spectrum of preequilibrium dileptons can dominate over that of other sources, in the intermediate invariant mass window. This suggests that dilepton production can provide a window to explore the properties of the preequilibrium stage after selecting the appropriate invariant mass range.  In the low invariant mass range, the flow results at intermediate centrality shown in Figs.~\ref{fig:v2_pt_centrality} and \ref{fig:v3_pt_centrality}, show a sensitivity to the chemical equilibrium effect. However, a proper assessment in this mass range will require the inclusion of dileptons from other hadronic sources. Indeed, the dileptons from reactions involving composite hadrons are also a very important source for low invariant masses dileptons \cite{Rapp:1999ej}.  Finally, adapting the dilepton rates to account for the various corrections discussed earlier in the text will be important for high precision future studies. Incorporating those many aspects in a comprehensive and detailed dilepton analysis can help us gain a deeper understanding of the dynamics of heavy ion collisions and the nature of the QCD phase diagram.

\acknowledgments{It is a pleasure to thank Greg Jackson for contributions that led to this work, for valuable discussions, and for a thorough reading of an earlier draft of this paper. This work was supported in part by the Natural Sciences and Engineering Research Council of Canada, and in part by US National Science Foundation (NSF) under grant number OAC-2004571. Computations were made on the B\'eluga super-computer system from McGill University, managed by Calcul Qu\'ebec and by the Digital Research Alliance of Canada.}

\appendix
\section{\label{sec:appendix}Drell-Yan dileptons from DYTurbo and Pythia6}
\begin{figure}[hb]
	\centering
	\includegraphics[width=0.9\linewidth]{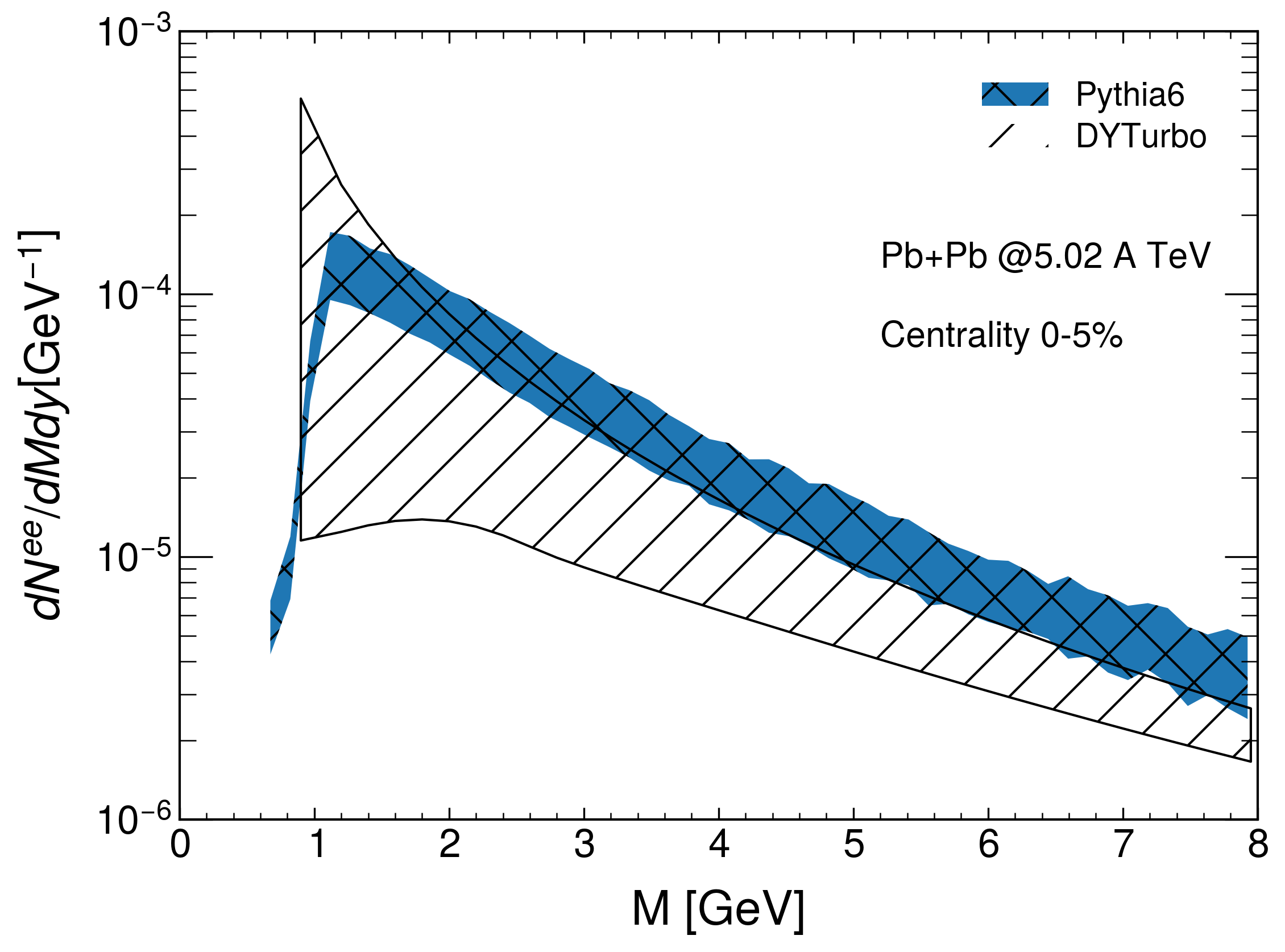}
	
 \caption{The dilepton production by the Drell-Yan process from DYTurbo and Pythia6 in the 0-5\% centrality class for Pb+Pb collision at the LHC.}
	\label{fig:DY_turbo_pythia}
\end{figure}
We compare the predictions of DYTurbo against those from Pythia6, for the production of dileptons via the Drell-Yan process. 
In Fig.~\ref{fig:DY_turbo_pythia}, the dilepton production from the Drell-Yan process estimated from the DYTurbo model and Pythia6 event generator in 0-5\% centrality Pb+Pb collisions at 5.02 TeV is shown. DYTurbo has the same set up as that used in Fig.~\ref{fig:dNdpt_M_0_5}. In the Pythia6 calculation, the parameter for the width of the Gaussian primordial \( k_T \) distribution is taken as 0.95 [PARP(9) = 0.95], the square of the momentum transfer equals the square of the dilepton pair mass  $Q^2 = M^2$ [MSTP(32) = 4], and it uses the default PDF sets CTEQ5L. The bands in Pythia6 result from varying the K factor, which multiplies the differential cross section for hard parton–parton processes to effectively account for higher order corrections, from 1.0 to 1.8 [PARP(31) = 1.0 - 1.8]. It is observed that dilepton production from the Drell-Yan process exhibits significant model uncertainty in both cases. Pythia6 predicts a higher dilepton yield compared to the DYTurbo calculation when the invariant mass $M>1.6$ GeV. But there is a clear region of overlap between both approach to the calculation of DY dileptons. Accurate estimates of the Drell-Yan process will significantly affect the extraction of dilepton production from the preequilibrium stage, and this highlights the importance of measuring the Drell-Yan process  at intermediate invariant masses.

\bibliography{references}

\end{document}

%% file: custom.tex
\def\be{\begin{equation}}
\def\ee{\end{equation}}
\def\bea{\begin{eqnarray}}
\def\eea{\end{eqnarray}}

\def\kompost{K\o{}MP\o{}ST }